\documentstyle[12pt,epsfig]{article}
\begin{document}

\noindent Stockholm\\
USITP 02\\
June 2002\\
Revised November 2002\\

\vspace{1cm}

\begin{center}

{\Large HOW TO MIX A DENSITY MATRIX}

\vspace{1cm}

{\large Ingemar Bengtsson}\footnote{Email address: ingemar@physto.se. 
Supported by VR.}

\

{\large \AA sa Ericsson}\footnote{Email address: asae@physto.se.}

\

{\sl Stockholm University, SCFAB\\
Fysikum\\
S-106 91 Stockholm, Sweden}

\vspace{8mm}

{\bf Abstract}

\end{center}

\vspace{5mm}

\noindent A given density matrix may be represented in many ways as a mixture of pure 
states. We show how any density matrix may be realized as a uniform 
ensemble. It has been conjectured that one may realize all probability 
distributions that are majorized by the vector of eigenvalues of the 
density matrix. We show that if the states in the ensemble are assumed 
to be distinct then it is not true, but a marginally weaker 
statement may still be true. 

\newpage

{\bf 1. Introduction}

\vspace{5mm}

\noindent A key property of quantum mechanics is that every mixed state, that is 
every non-pure density matrix, can be written as an ensemble of pure states in 
many ways. There exists a well known characterization of this property, apparently 
first published by Schr\"odinger \cite{Erwin} \cite{HJW}:

\

\noindent {\bf Theorem 1}: A density matrix ${\rho}$ having the diagonal form 

\begin{equation} {\rho} = \sum_{i = 1}^M{\lambda}_i|e_i\rangle \langle e_i| 
\label{1} \end{equation}

\noindent can be written in the form 

\begin{equation} {\rho} = \sum_{i = 1}^Np_i|{\psi}_i\rangle \langle {\psi}_i| \ , 
\hspace{8mm} \sum_{i = 1}^Np_i = 1 \label{2} \end{equation}

\noindent if and only if there exists a unitary $N \times N$ matrix $U$ 
such that 

\begin{equation} |{\psi}_i\rangle = \frac{1}{\sqrt{p}_i}\sum_{j = 1}^MU_{ij}
\sqrt{{\lambda}_j}|e_j\rangle \ . \label{3} \end{equation}

\noindent Here all states are normalized to unit length but may not be 
orthogonal to each other.

\

\noindent Observe that the matrix $U$ does not act on the Hilbert space 
but on vectors whose components are state vectors, and also that we may well 
have $N > M$. But only the first $M$ columns of $U$ appear in 
the equation---the remaining $N - M$ columns are just added in order to allow 
us to refer to the matrix $U$ as a unitary matrix. What the theorem 
basically tells us is that the pure states $|{\psi}_i\rangle $ that make up 
an ensemble are linearly dependent on the $M$ vectors $|e_i\rangle $ that 
make up the so called ``eigenensemble''. Moreover an arbitrary state 
in that linear span can be included. For definiteness we assume from now 
on that all density matrices have rank $M$ so that we 
consider ensembles of $N$ pure states in an $M$ dimensional Hilbert space. 

One can say a bit more. Recall that there is a notion called ``majorization'' 
that provides a natural partial preordering of probability distributions 
\cite{Ando}. To be precise assume that (if necessary) the 
eigenvalue vector $\vec{\lambda}$ 
has been extended with zeroes until it has the same number $N$ of components 
as $\vec{p}$, and also that the entries in the probability vectors have 
been arranged in decreasing order. (In the sequel these assumptions will often 
be made tacitly.)  By definition 
the distribution $\vec{p}$ is majorized by the distribution 
$\vec{\lambda}$, written $\vec{p} \prec \vec{\lambda}$, if and only if 

\begin{equation} \sum_{i = 1}^k p_i \leq \sum_{i = 1}^k {\lambda}_i 
\label{4} \end{equation}

\noindent for all $k < N$. In colloquial terms, the probability distribution 
$\vec{p}$ is ``more even'' than the distribution $\vec{\lambda}$. Now it is an easy 
consequence of Theorem 1 that the probability vector that appears there is 
given by

\begin{equation} p_i = \sum_{j=1}^MB_{ij}{\lambda}_j \ ; \hspace{8mm} B_{ij} = 
|U_{ij}|^2 \ . \label{5} \end{equation}

\noindent The matrix $B$ is bistochastic (all its matrix elements are positive 
and the sum of each row and each column is unity). This follows because by 
construction it is unistochastic, that is each matrix element is the absolute 
value squared of the corresponding element of a unitary matrix. All unistochastic 
matrices are bistochastic, but the converse is not true. One can now show:

\

\noindent {\bf Theorem 2}: Given a probability vector $\vec{p}$ there exists a 
set of pure states $|{\psi}\rangle $ such that eq. (\ref{2}) holds if and only if 
$\vec{p} \prec \vec{\lambda}$, where $\vec{\lambda}$ is the eigenvalue vector.

\

\noindent In one direction this was shown by Uhlmann \cite{Uhlmann}: If such a 
decomposition exists then $\vec{p} \prec \vec{\lambda}$ because all vectors that 
can be reached from a given vector with a bistochastic matrix are majorized by 
the given vector \cite{Ando}. The converse was shown by Nielsen, who gave an 
algorithm for constructing the states $|{\psi}_i\rangle $ given the vector 
$\vec{p} \prec \vec{\lambda}$ \cite{Nielsen}. 
We will return to his construction below. 

Why are these facts of interest? For one thing the components of $\vec{p}$ can 
arise as the squares of the coefficients in a Schmidt decomposition of a 
bipartite entangled state (and the density matrix then appears as the state of 
a subsystem). These theorems then give insight into the different representations 
that entangled states can be given. In particular Nielsen uses this insight to 
obtain a new protocol for the conversion of one entangled state to another by 
means of local operations and classical communication (LOCC) \cite{Nielsen}. As a 
general remark increasing ability to manipulate quantum states in the laboratory 
requires increasing precision in our understanding of 
how they can be represented mathematically. 

Now we can make a more precise statement which has not been proved:

\

\noindent {\bf Conjecture 1}: Given any probability vector $\vec{p}$ majorized by 
the eigenvalue vector $\vec{\lambda}$ there exists a set of distinct pure states 
$|\psi_i \rangle $ such that eq. (\ref{2}) holds. 

\

\noindent There is no guarantee that the algorithm offered by Nielsen leads to 
an ensemble of distinct states. In fact, as we will show, in general it does 
not---indeed Conjecture 1 is false. 

Our purpose is to formulate a new conjecture along the same lines 
that has a chance of being true. We begin in 
section 2 with a geometrical proof of a weak 
form of Conjecture 1, namely that any non-pure density matrix 
can be obtained as a uniform ensemble of pure states. We also explain 
why Conjecture 1 is in fact false. In section 3 we 
provide a review for physicists of the theory of majorization and 
bistochastic matrices. In section 4 we analyze the counterexamples to 
Conjecture 1 and collect some evidence for Conjecture 2, which will 
be a slight modification of the original. We fail to prove it though. 
Our conclusions are summarized in section 5. 

\vspace{1cm}

{\bf 2. Uniform ensembles}

\vspace{5mm}

\noindent We want to construct a uniform ensemble (with all the $p_i$ equal 
to $1/N$) for an arbitrary quantum state. Provided that the state is not 
pure such an ensemble can be constructed 
with very little ado. Let  ${\rho} = \mbox{diag}({\lambda}_1, {\lambda}_2, 
\ ... \ , {\lambda}_M)$. Choose a pure state vector whose entries are the square 
roots of the eigenvalues of ${\rho}$. Then form the one parameter family of 
state vectors 

\begin{equation} Z^{\alpha}({\tau}) = \left( \begin{array}{ccc} e^{in_1{\tau}} & 
0 & 0 \\ 0 & e^{in_2{\tau}} & 0 \\ 0 & 0 & e^{in_3{\tau}} \end{array} 
\right) \left( \begin{array}{c} \sqrt{{\lambda}_1} \\ \sqrt{{\lambda}_2} \\ 
\sqrt{{\lambda}_3} \end{array} \right) \ . \end{equation}

\noindent Here we choose $M = 3$ for illustrative purposes and the notation anticipates 
the fact that we will choose the $n_i$ to be integers. Rewrite these state vectors 
as projectors, 

\begin{equation} Z^{\alpha}({\tau})
\bar{Z}_{\beta}({\tau}) = \left( \begin{array}{ccc} {\lambda}_1 & 
\sqrt{{\lambda}_1{\lambda}_2}e^{in_{12}{\tau}} & \sqrt{{\lambda}_2{\lambda}_3}
e^{in_{13}{\tau}} \\ \sqrt{{\lambda}_2{\lambda}_1}e^{in_{21}{\tau}} & {\lambda}_2 & 
\sqrt{{\lambda}_2{\lambda}_3}e^{in_{23}{\tau}} \\ 
\sqrt{{\lambda}_3{\lambda}_1}e^{in_{31}{\tau}} & 
\sqrt{{\lambda}_3{\lambda}_2}e^{in_{32}{\tau}} & {\lambda}_3 \end{array} 
\right) \ . \end{equation}

\noindent where $n_{ij} \equiv n_i - n_j$. Form a uniform ensemble of pure 
states by 

\begin{equation} {\rho}' = \frac{1}{2{\pi}} \int_0^{2{\pi}}d{\tau}Z^{\alpha}({\tau})
\bar{Z}_{\beta}({\tau}) \ . \end{equation}

\begin{figure}
        \centerline{ \hbox{
                \epsfig{figure=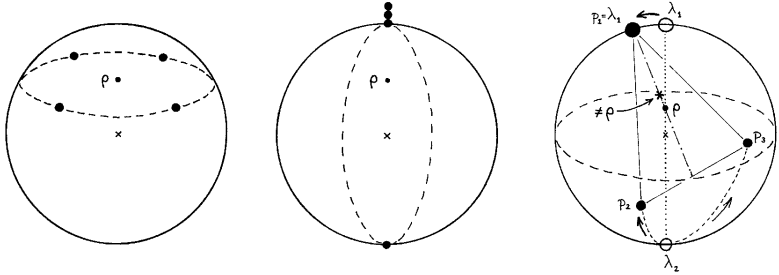,width=11cm}}}
        \caption{{\small To the left we see a uniform distribution with 
$N = 4$ for the 
density matrix ${\rho} = \mbox{diag}(3/4, 1/4)$ in the Bloch ball and in 
the middle the degenerate uniform distribution that one would obtain from 
Nielsen's procedure. By studying the rightmost picture 
one can convince oneself that if $p_1 = {\lambda}_1$ then any ensemble 
with $N > 2$ pure states must be degenerate.}}
        \label{fig:ett}
\end{figure}

\noindent Clearly if we choose the $n_i$ such that all the $n_{ij}$ are 
non-zero integers we will get ${\rho}' = {\rho}$, as was our aim. In geometrical 
terms, what we are doing is to represent our density matrix as a uniform 
distribution on a suitable closed 
Killing line, that is a flowline of a unitary transformation that leaves the 
original density matrix invariant. It is also clear that we can get a finite 
distribution by placing $N$ points on the closed curve parametrized by ${\tau}$, 
using the roots of unity. Finally it is clear that the argument works for all $M$. 
We illustrate the case $M = 2$ in fig. \ref{fig:ett}. Note that if we 
regard the space of 
density matrices equipped with the Hilbert-Schmidt metric as a subset of a flat 
Euclidean space then the closed curve is not a circle in that flat space 
except when $M = 2$, but then this is not needed for the argument. 

%\vspace{3cm}
%\noindent {\small 
%}
%\vspace{5mm}

Clearly there are many ways to realize a uniform ensemble. Nielsen \cite{Nielsen} 
provides a different procedure that relies on the theorem by Horn \cite{Horn}, 
discussed in the next section. Horn's theorem tells us how to construct a 
matrix $U$ that obeys eq. (\ref{5}) whenever $\vec{p}$ is majorized by 
$\vec{\lambda}$. This matrix is then used in eq. (\ref{3}). But 
there is no guarantee that the states $|{\psi}_i\rangle $  are distinct. 
Explicit calculation shows that generically they are not, except when 
$N = M$. What this shows is 
that Theorem 1 must be used with some care. While the rows of a unitary 
matrix are never equal, it may still be true that the first $M$ components 
of a pair of rows in a unitary $N \times N$ matrix coincide. If this happens 
two of the pure states in the decomposition (\ref{2}) will coincide too, and 
the ensemble will in fact not be uniform (see fig. \ref{fig:ett}). 
An analogous difficulty 
affects non-uniform ensembles as well. More details will be provided in section 4, 
once we have sketched some relevant background.

Further inspection of the Bloch ball reveals that Conjecture 1 cannot be 
true in general. It obviously fails for pure states. A more interesting 
counterexample is the following: Let $({\lambda}_1, {\lambda}_2) = (1/2, 1/2)$. 
This is the maximally mixed state. Let $(p_1, p_2, p_3) = (\frac{1}{2}, 
\frac{1}{4}, \frac{1}{4})$. Clearly 
$\vec{p} \prec \vec{\lambda}$. But it is geometrically evident that an 
ensemble of pure 
states with these $p_i$ as probabilities cannot give the maximally mixed state: Our 
three states define a plane through the Bloch ball that has to go through the 
center of the ball (where the maximally mixed 
state sits). The convex sum of the two states with probability $1/4$ lies inside 
the ball and the density matrix must lie on the straight line between that point 
and that of the state with probability $1/2$. But then the density matrix cannot 
lie in the center of the ball as stated. An extension of this argument shows 
that when $M = 2$ it is always impossible to realize a non-degenerate 
ensemble with $p_1 = {\lambda}_1$ and $N > 2$. See fig. 1.

\vspace{1cm}

{\bf 3. Majorization and bistochastic matrices.}

\vspace{5mm}

\noindent To bring the issues into focus a review of the mathematical 
background is called for. Majorization, as defined in the introduction, 
provides a natural partial preordering of vectors, and in particular of 
discrete probability distributions (positive vectors with trace norm 
equal to one). It is a preordering because $\vec{p} \prec \vec{q}$ and 
$\vec{q} \prec \vec{p}$ does not imply $\vec{p} = \vec{q}$, only that 
the vector $\vec{q}$ is obtained by permuting the components of $\vec{p}$. 
The notion is important in many contexts, ranging from economics to 
LOCC (Local Operations and Classical Communication) of entangled states 
in quantum mechanics \cite{Nielsen2}. 

\begin{figure}
        \centerline{ \hbox{
                \epsfig{figure=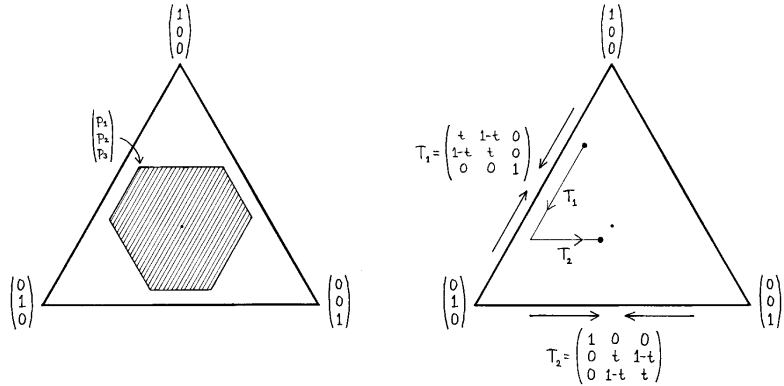,width=11cm}}}
        \caption{{\small The probability simplex for $N = 3$. To the left we 
see the set of vectors majorized by a given vector sitting in the corner 
of the shaded polytope. To the right we see how to get from one probability 
vector to another by means of two $T$--transforms.}}
        \label{fig:tva}
\end{figure}

The set of probability vectors is a convex simplex, and the set of such 
vectors that are majorized by a given vector forms a convex polytope with 
its corners at the $N!$ vectors obtained by permuting the $N$ components of 
the given vector. It is helpful to keep the simplest example in mind. Let 
the number of components be $N = 3$. Then the set forms a triangle, and 
the set of vectors majorized by a given vector is easily recognized 
(see fig. \ref{fig:tva}).

%\vspace{3cm
%\noindent {\small }
%\vspace{5mm} 

A basic fact about majorization is a theorem due to Hardy, Littlewood 
and P\'olya, that states that $\vec{p} \prec \vec{q}$ if and only if 
there exists a bistochastic matrix $B$ such that $\vec{p} = B\vec{q}$. 
A stochastic matrix is a matrix with non--negative entries such that 
the elements in each column sum to unity, which means that the matrix 
transforms probability vectors to probability vectors. It is 
bistochastic if also its rows sum to unity, which means that 
the uniform distribution $\vec{e} = \frac{1}{N}(1, 1, \ ... \ , 1)$ 
is a fixed point of the map. According to Birkhoff's theorem the 
space of bistochastic $N$ by $N$ matrices is an $(N-1)^2$ dimensional 
convex polytope with the $N!$ permutation matrices making up its 
corners. In the center of the polytope we find the van der Waerden 
matrix $B_*$ all of whose entries are equal to $1/N$. 

Some special cases of bistochastic matrices will be of 
interest below. A $T$--transform is a bistochastic matrix that acts 
non-trivially only on two entries of the vectors. By means 
of permutations it can therefore be brought to the form

\begin{equation} T = \left( \begin{array}{ccccc} t & 1-t & 0 & \cdots & 0 \\
1-t & t & 0 & \cdots & 0 \\ 0 & 0 & 1 & \cdots & 0 \\ \vdots & \vdots 
& \vdots & \ddots & \vdots \\ 0 & 0 & 0 & \cdots & 1 \end{array} \right) 
\ , \hspace{5mm} 0 \leq t \leq 1 \ .  \end{equation}

\noindent Given two vectors $\vec{p} \prec \vec{q}$ there always 
exists a sequence of $N - 1$ $T$--transforms such that $\vec{p} = 
T_{N-1} \ ... \ T_1\vec{q}$. On the other hand (except when $N = 2$) 
it is not true that every bistochastic matrix can be written as a 
sequence of $T$--transforms. Fig. \ref{fig:tva} shows how 
$T$--transforms act when $N = 3$. 

\begin{figure}
        \centerline{ \hbox{
                \epsfig{figure=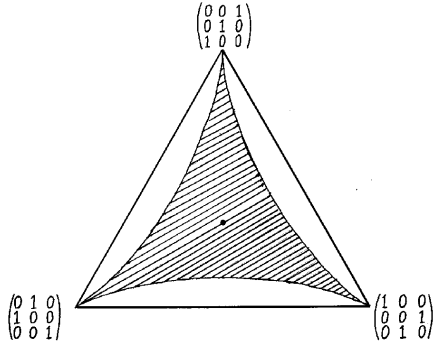,width=6cm}}}
        \caption{{\small A two dimensional slice through Birkhoff's 
polytope; the shaded region consists of unistochastic matrices and the 
dot in the center represents the van der Waerden matrix. Some 
interesting observations are made about it in ref. \cite{Karol}, 
where a closely related picture is drawn.}}
        \label{fig:tre}
\end{figure}

Unistochastic matrices were defined in the introduction. An 
orthostochastic matrix is a special case of that where the matrix 
elements of the bistochastic matrix are given by squares of the 
corresponding element of an orthogonal matrix. Horn's theorem 
\cite{Horn} states that given $\vec{p} \prec \vec{q}$ one can 
always find an orthostochastic matrix $B$ such that $\vec{p} = 
B\vec{q}$. The proof is by induction and actually gives a 
construction of $B$ as a sequence of $N-1$ $T$--transforms. 
The set of unistochastic matrices form a compact connected subset of 
the set of bistochastic matrices. When $N > 2$ not all bistochastic 
matrices are unistochastic (see fig. \ref{fig:tre}). The van der 
Waerden matrix is unistochastic. It will be of 
interest below to know that sequences of $T$--transforms are always 
unistochastic when $N = 3$, but that this is not so when $N > 3$ 
\cite{Poon}. For a more extensive discussion of unistochastic 
matrices we recommend reference \cite{Karol}.

\vspace{1cm}

{\bf 4. Conjectures.}

\vspace{5mm}

\noindent We can now return to the question of precisely which 
probability vectors $\vec{p}$ that can occur in non-degenerate 
ensembles for a density matrix with eigenvalue vector $\vec{\lambda}$. 
We know that $\vec{p} \prec \vec{\lambda}$. Nielsen's idea 
\cite{Nielsen} is to rely on Horn's theorem to provide an orthostochastic 
matrix $B$ connecting the two vectors. The catch---from our point of 
view---is that the resulting ensemble will be degenerate in generic cases. 

In section 2 we pointed out a class of counterexamples to Conjecture 
1 for two dimensional Hilbert spaces. We can now see in a different way how they 
arise for arbitrary dimension $N$. Let 

\begin{equation} p_1 + \ ... \ + p_{k-1} = {\lambda}_1 + 
\ ... \ + {\lambda}_{k-1} \label{11} \end{equation}

\noindent and assume ${\lambda}_{k-1} > {\lambda}_k$. 
It is then easy to show that any bistochastic matrix connecting the 
two vectors must take the block diagonal form 

\begin{equation} B = \left( \begin{array}{cc} D_1 & 0 \\ 0 & D_2 
\end{array} \right) \ , \end{equation}

\noindent where $D_1$ and $D_2$ are bistochastic matrices in themselves. 
$D_1$ is a $(k-1)\times (k-1)$ matrix. If $k = M$ then the form of 
$B$ means that only one column of $D_2$ is actually used in constructing 
no less than $N - M + 1$ of the pure states in eq. (\ref{3}), so that 
all these state vectors are contained in a one dimensional subspace. Hence the 
ensemble must have at least that degree of degeneracy, for essentially 
the same reason that a pure density matrix leads to a totally degenerate 
ensemble. It is tempting to guess that this is the only kind of 
counterexample to Conjecture~1 that can arise. Note however that 
the geometric argument in section 2 was actually a little 
more general as far as the case $M = 2$ is concerned, and excludes 
also the case ${\lambda}_{M-1} = {\lambda}_M$. 

When $k < M$ in eq. (\ref{11}) there is no obvious reason why 
the block diagonal form of $B$ must lead to degeneracies, and in 
fact this is not so in the (few) examples that we checked. In the 
concluding section we will conjecture that we have already found 
all the necessary additional restrictions that are missing from 
Conjecture 1. 

Let us approach the problem from another direction. Given $\vec{p} 
\prec \vec{\lambda}$ can we find an algorithm for how to construct a 
unistochastic matrix $B$ such that $\vec{p} = B\vec{\lambda}$ and such 
that the corresponding unitary matrix leads to a non-degenerate ensemble? 
To begin with let us assume that $\vec{p} = \vec{e}$ and let us ask for 
a sequence of $T$--transforms that produces the ``natural'' uniform 
ensemble presented in section 2. An algorithm that does this is to 
first apply a $T$--transform $T_1$ with $t = 1/2$ to the first two entries 
in $\vec{p}$, then a $T$-transform $T_2$ with $t = 1/2$ to the second 
and third entries, and so on until $T_{N-1}$ sets the last 
two components of the vector equal. Then we repeat the procedure an 
infinite number of times. We get 

\begin{equation} \lim_{n \rightarrow \infty }(T_{N-1}T_{N-2} \ 
... \ T_1)^n = \frac{1}{N} \left( \begin{array}{ccccc} 1 & 1 & 1 
& \cdots & 1 \\ 1 & 1 & 1 & \cdots & 1 \\ 1 & 1 & 1 & \cdots & 1 \\ 
\vdots & \vdots & \vdots & \ddots & \vdots \\ 1 & 1 & 1 & \cdots 
& 1 \end{array} \right) \ . 
\end{equation}

\noindent This is the van der Waerden matrix $B_*$ which is unistochastic and 
when used in Schr\"odinger's theorem does in fact lead to the natural uniform ensemble 
from section 2. To see that the sequence converges to $B_*$ we simply 
observe that the $T$--transforms do not depend on the vector $\vec{p}$ 
that we start out with. Therefore we can read off the columns of $B_*$ by 
seeing how it acts on the corners of the probability simplex, that is the 
vectors $(1, 0, \ ... \ , 0)$ and so on. 

Clearly this algorithm can be generalized to arbitrary $\vec{p}$ and 
$\vec{\lambda}$. The key idea is to choose, at each step, a $T$--transform 
that ensures 
the equality $p_{k}/p_{k-1} = {\lambda}_k/{\lambda}_{k-1}$. Typically 
this will again converge to a definite bistochastic matrix $B$ in an 
infinite number of steps, this time because the individual $T$--transforms 
approach the unit matrix. In the examples that we checked (mostly for 
three-by-three matrices) it does produce a non-degenerate ensemble 
except for the counterexamples we already have. We therefore have a 
candidate for a constructive algorithm with the desired properties.

Unfortunately we do not know if the candidate is good enough. We do not 
know that it always results in a unistochastic matrix, let alone a non-degenerate 
ensemble. Already the first question becomes non-trivial when $N >3$, as 
noted in section 3. What we do know is that a sequence of 
$T$--transforms always meets the ``chain-links'' conditions from ref. 
\cite{Karol2} (see also \cite{Karol}). These 
are necessary but not sufficient conditions that a bistochastic matrix 
is unistochastic. The idea is as follows: Take a bistochastic matrix 

\begin{equation} B = \left( \begin{array}{ccc} a_1 & b_1 & c_1 \\ 
a_2 & b_2 & c_2 \\ a_3 & b_3 & c_3 \end{array}\right) \ . \end{equation}

\noindent Form the ``links'' $L_i = \sqrt{a_ib_i}$. If $B$ is unistochastic 
it must be true that 

\begin{equation} L_1 \leq L_2 + L_3 \ , \hspace{3mm}L_2 \leq L_3 + L_1 
\hspace{3mm} \mbox{and} \hspace{3mm} L_3 \leq L_1 + L_2 \ . \end{equation}

\noindent These are called the chain-links conditions (stated in terms of 
columns) because they make it possible to form a ``chain'' 
(in this case a triangle) out of the links. This in turn ensures that a set 
of phases ${\mu}_1$, ${\mu}_2$ can be found such that the matrix

\begin{equation} \left( \begin{array}{lll} \sqrt{a_1} & \sqrt{b_1} & \cdot \\ 
\sqrt{a_2} & \sqrt{b_2}e^{i{\mu}_1} & \cdot \\ \sqrt{a_3} & 
\sqrt{b_3}e^{i{\mu}_2} & \cdot \end{array} \right) \end{equation}

\noindent is unitary. (It is unnecessary to check the last column.) For 
three by three matrices the chain-links conditions are sufficient. For 
$N$ by $N$ matrices with $N > 3$ the story becomes more complicated. The 
chain-links conditions still state that no one of the lengths (constructed 
analogously) can be larger than the sum of all the others. When one 
tries to construct the unitary matrix the number of equations to solve is 
the same as the number of phases available, but it can happen that 
the equations have no solution. Therefore the chain-links conditions 
are necessary but not sufficient when $N > 3$. 

\

\noindent {\bf Lemma:} A sequence of $T$--transforms always obeys the 
chain-links conditions. 

\

\noindent Sketch of proof: 
A single $T$--transform is unistochastic. Consider any sequence of  
$T$--transforms. Suppose that the first $n$ of these form a matrix $B$ 
that obeys the chain-links conditions. It is enough to prove that this 
implies that $TB$ obeys the chain-links conditions, 
where $T$ is any $T$--transform. (Since the set of matrices that obey 
these conditions is 
compact, infinite sequences pose no particular problem.) Consider therefore 

\begin{equation} T(t)B = \left( \begin{array}{ccc} 1 & 0 & 0 \\ 
0 & t & 1 - t \\ 0 & 1 - t & t \end{array}\right)  
\left( \begin{array}{ccc} a_1 & b_1 & \cdot \\ 
a_2 & b_2 & \cdot \\ a_3 & b_3 
& \cdot \end{array}\right) \ , \hspace{5mm} 0 \leq t 
\leq 1 \ . \end{equation}

\noindent With the links $L_i(t)$ formed from the matrix $T(t)B$ as 
indicated above, we must show that 

\begin{equation} L_1(1) \leq L_2(1) + L_3(1) 
\hspace{5mm} \Rightarrow \hspace{5mm} L_1(t) \leq L_2(t) + L_3(t) 
\ . \end{equation}

\noindent (Note that $T(1)B = B$ which obeys the conditions by assumption.) 
Since $L_1(t)$ is constant it is enough to verify that  
the function $L_2(t) + L_3(t)$ is larger than $L_2(1) + 
L_3(t)$ everywhere in the interval. We observe that this function 
is symmetric around $t = 1/2$ and a straightforward calculation 
verifies that it assumes its only extremum there, and that this 
extremum is a maximum. The proof that the other 
chain-links conditions hold is similar. Extension of the proof to larger 
matrices and to the chain-links condition stated in terms of rows rather 
than columns is also straightforward.

For three-by-three matrices this establishes the result of ref. \cite{Poon}, 
that any sequence of $T$--transforms is unistochastic. For larger matrices 
the chain-links condition is necessary but not sufficient for that and there 
do exist sequences of $T$--transforms that are not unistochastic. It remains 
possible that the particular kind of sequence that we propose as an algorithm 
to realize an ensemble with probability vector $\vec{p}$ always results 
in unistochastic matrices but we have failed to prove this. Assuming that 
this can be done we would still have to prove that the ensemble that results 
from the unitary matrix so constructed is non-degenerate for all allowed 
probability vectors. This appears to be significantly more difficult.

\vspace{1cm}

{\bf 5. Conclusions.}

\vspace{5mm}

\noindent We have studied the question of precisely 
what kind of discrete probability distributions that can appear in an ensemble 
of pure states that describe a given density matrix. Evidently our 
main conclusion is that this question has more facets to it than one might 
suspect based on earlier literature \cite{Nielsen}. We think 
that one answer is the following: 

\

\noindent {\bf Conjecture 2}: Given any probability vector $\vec{p}$ majorized by 
the eigenvalue vector $\vec{\lambda}$. Assume that the density matrix is not 
pure. Then there exists a set of distinct pure states 
$|\psi_i \rangle $ such that eq. (\ref{2}) holds if and only if 
$p_1 + \ ... \ + p_{M-1} \neq {\lambda}_1 + \ ... \ + {\lambda}_{M-1}$. 

\

\noindent The ``only if'' part of the statement is proved for $M = 2$. For 
$M > 2$ the case ${\lambda}_{M-1} = {\lambda}_M$ may need separate 
attention, otherwise the ``only if'' part is again proved. The ``if'' part 
is only weakly supported by arguments and examples. 

We also made a suggestion for how, given a $\vec{p}$ consistent with 
Conjecture 2, one might go about to 
construct such an ensemble by means of a sequence of $T$--transforms, but 
this suggestion is very weakly supported. Essentially the only 
argument is that the procedure does give the geometrically natural uniform 
ensemble when $\vec{p} = \vec{e}$.
 
\vspace{1cm}

{\bf Acknowledgement:}

\vspace{5mm}

\noindent We thank G\"oran Lindblad, Wojciech S{\l}omczy\'nski 
and Karol \.Zyczkowski for discussions and help; also an anonymous 
referee for comments.

%\newpage

\begin{thebibliography}{99}

\bibitem{Erwin} E. Schr\"odinger, Proc. Camb. Phil. Soc. \underline{32} (1936) 446.

\bibitem{HJW} L. P. Hughston, R. Josza and W. K. Wootters, Phys. Lett. 
\underline{A183} (1993) 14.

\bibitem{Ando} T. Ando, Lin. Alg. Appl. \underline{118} (1989) 163.

\bibitem{Uhlmann} A. Uhlmann, Rep. Math. Phys. \underline{1} (1970) 147. 

\bibitem{Nielsen} M. A. Nielsen, Phys. Rev. \underline{A62} (2000) 052308.

\bibitem{Horn} A. Horn, Amer. J. Math. \underline{76} (1954) 620.

\bibitem{Nielsen2} M. A. Nielsen, Phys. Rev. Lett. \underline{83} (1999) 436.

\bibitem{Poon} Y.-T. Poon and N.-K. Tsing, Lin. Multilin. Alg. 
\underline{21} (1987) 253.

\bibitem{Karol} K. \.Zyczkowski, M. Ku\'s, W. S{\l}omczy\'nski and H.-J. 
Sommers, Random unistochastic matrices, arXiv preprint nlin.CD/
0112036.

\bibitem{Karol2} P. Pako\'nski, K. \.Zyczkowski and M. Ku\'s, J. Phys. 
\underline{A34} (2001) 9303.

\end{thebibliography}
\end{document}